\documentstyle[prl,aps,twocolumn]{revtex} 
\newcommand{\avg}[1]{\langle{#1}\rangle}
\newcommand{\req}[1]{(\ref{#1})}
\newcommand{\beq}{\begin{equation}}
\newcommand{\beqar}{\begin{eqnarray}}
\newcommand{\eeqar}{\end{eqnarray}}
\newcommand{\beqars}{\begin{eqnarray*}}
\newcommand{\eeqars}{\end{eqnarray*}}
\newcommand{\eeq}{\end{equation}}

\begin{document}
\baselineskip=0.97\baselineskip
{\bf Comment on: Role of Intermittency in Urban Development: 
A Model of Large-Scale City Formation}\\
In Ref. \cite{zm} a model for large-scale city 
formation was proposed. It is based on the 
discreet dynamics (eqs. (1,2) of \cite{zm})
\beq
n_i(t+1)=(1-\alpha)n_i(t)f_i(t)+\frac{\alpha}{4}
\sum_{j~nn~i} n_j(t)f_j(t), 
\label{eq1}
\eeq
where $n_i(t)$ is the population at site
$i$ of a square lattice, and $f_i(t)\ge 0$ 
are random multiplication factors drawn independently 
from a distribution $P(f)$ with 
the property $\avg{f}=1$.
The authors claim that such a multiplicative
process with diffusion gives rise to a stationary state 
with a power law distribution of city sizes $n_i$
$P(n)\sim n^{-\tau}$, where $\tau \simeq 2$ in their simulations 
is in agreement with empirical observations or real city 
size distribution.

We demonstrate here that {\em i)} Eq. \req{eq1}, with
$\avg{f}=1$, does not lead to a stationary state;
{\em ii)} the probability distribution 
of $n_i$ does not have a power-law tail; {\em iii)}
both stationarity and a power-law tails of the 
distribution are recovered if a 
constant source term is added to the RHS of the Eq. \req{eq1}, 
but then the uniqueness of the exponent $\tau=2$ is lost.

We have performed a  numerical simulation of 
the dynamical rules \req{eq1} 
and found that, contrary to the 
claims of Ref.  \cite{zm}, these dynamical rules 
{\it do not lead to a stationary 
state}. Instead, the total population $N(t)=\sum_i n_i(t)$ decays 
to zero exponentially in time. 
For example, for $\alpha=0.25$, and $f$ being equal to
$2$ or $0$ with equal probabilities, we found that 
after $1000$ time steps $N(t)$ is of order $10^{-15}$
in clear contradiction with the results of Ref. \cite{zm} 
for the same set of parameters.
This is not a finite size effect (we found the same 
rate of decay for sizes in the range $64^2$--$200^2$). 
This inconsistency suggests that there is some ingredient
in the model simulated by Zanette and Manrubia, which was not 
reported in their paper. 

To understand better the reasons for this exponential decay
let us consider first the case $\alpha=0$, where 
the population $n_i(t)$ at each site undergoes a random 
multiplicative process 
$n_i(t+1)=n_i(t)f_i(t)$
and is uncoupled from other sites. 
For $\avg{f}=1$ the expectation value $\avg{n_i(t)}$ does
not change in time. It is known, however, that 
in such process, $\avg{n_i(t)}$
is dominated by extremely rare events, when $n_i$
is exponentially large in $t$. 
By the virtue of Central Limit Theorem, in any {\em typical} 
realization, $n_i(t)\simeq e^{\avg{ \ln f} t}$.
One can show that for any $P(f)$ such that 
$\avg{f}=1$, one has $\avg{\ln f}<0$, 
so that the typical $n_i (t)$ 
vanishes exponentially for $t\to\infty$. 
The diffusion ($\alpha>0$) alone cannot reverse
this typical decay. Indeed, the total population 
also undergoes a multiplicative
process: $N(t+1)=F(t)N(t)$, where $F=\sum_i f_i 
n_i/N$ is the ``population'' average of $f_i$. 
Again $\avg{F}=1$ implies that $\avg{\ln F}<0$,
so that typically $N(t)\sim e^{\avg{\ln F} t}\to 0$ 
for $t\to\infty$. 

Note that Eq. \req{eq1} is the equation for the 
partition function $\sum _i n_i(t)$ of a directed polymer
of length $t$ in a three-dimensional 
random media \cite{hz}, and $-\avg{\ln F}$ is the 
polymer's free energy per unit length.
For this problem it is known\cite{hz} that 
$P(n)$ is not a power law but rather a
log-stretched-exponential law.

On the other hand, if a positive constant 
is added to the right hand side of Eq. 
\req{eq1}, it describes a multiplicative noise 
process with a lower wall. The purpose 
of this extra source term is to prevent $n_i(t)$ from
becoming too small, while for large $n_i$ its influence can be 
safely neglected. Such rules dynamics have received much attention 
recently both for $\alpha>0$ \cite{mh}, and $\alpha=0$ \cite{cs}.
For $\alpha=0$ it was shown \cite{cs}
that a stationary state exists, provided $\avg{\ln f}<0$.
The distribution of $n$ has a power law tail $n^{-\tau}$ with
$\tau$ determined by $\avg{f^{\tau-1}}=1$. 
Clearly $\avg{f}=1$ implies $\tau=2$, 
but the condition $\avg{f}=1$ is no longer necessary 
for the existence of a stationary state, and cannot be
justified on these grounds.
For $\alpha>0$ we found that the system is
stationary (it is in the ``pinned''
phase \cite{mh}) and $P(n)$ still has a power 
law tail with $\tau=2$.
This can be justified by the fact that the correlation length
$\xi(\alpha)$ is finite away from the depinning 
transition \cite{mh}.
Therefore, a large system can be divided into 
many virtually uncoupled blocks, whose total population
$\tilde n_x=\sum_{|i-x|<\xi(\alpha)} n_i$
undergoes a multiplicative dynamics 
$\tilde n_x(t+1)=\tilde f_x(t)\tilde n_x(t)$ 
with $\avg{\tilde f}=1$. The last condition
ensures that $\tau=2$ for any $\alpha$. 
If $\avg{f}\not = 1$ instead $\tau$ depends 
on $\alpha$.

In short, the results of ref. \cite{zm} are inconsistent with
their rules of dynamics \req{eq1}, but compatible 
with the behavior of models with a lower wall \cite{mh}.
For such models a stationary state does not require 
$\avg{f}=1$, and any power law exponent 
$\tau>1$ is feasible.
\\
~\\
Matteo Marsili$^{(1,2)}$, Sergei Maslov$^{(3)}$,\\ 
and Yi-Cheng Zhang$^{(1)}$\\
(1) Institut de Physique Th\'eorique, 
Universit\'e de Fribourg 
P\'erolles, Fribourg, CH-1700, Switzerland.\\
(2) International School for Advanced Studies (SISSA)
and INFM unit, V. Beirut 4, I-34014 Trieste.\\
(3) Department of Physics, Brookhaven National Laboratory 
Upton, New York 11973.

\end{document}